\documentclass{article}

\usepackage[utf8]{inputenc}
\usepackage{graphicx}
\usepackage{placeins}
\usepackage{hyperref}
\usepackage[style=numeric, sorting=none, maxnames=99]{biblatex}
\usepackage{latexsym,amsmath,amssymb,amsfonts}
\usepackage{authblk}
\usepackage{setspace}

\newcommand{\ID}{M}
\newcommand{\SL}{V}
\newcommand{\DL}{K}
\newcommand{\AD}{T}
\newcommand{\HS}{D}
\renewcommand{\d}{\,\mathrm{d}}

\newcommand{\dy}{\d \mathbf{y}}
\newcommand{\Rthree}{\mathbb{R}^3}
\newcommand{\normal}{\hat{\mathbf{n}}}
\newcommand{\bonedomain}{\Omega_\mathrm{bone}}
\newcommand{\exteriordomain}{\Omega_\mathrm{ext}}

\title{Full-Wave Modeling of Transcranial Ultrasound using Volume-Surface Integral Equations and CT-Derived Heterogeneous Skull Data\footnote{© 2026. This manuscript version is made available under the CC-BY-NC-ND license. This manuscript is published in \emph{Ultrasonics} in final form at \url{https://doi.org/10.1016/j.ultras.2026.107954}.}}

\author[1]{Alberto Almuna-Morales}
\author[1]{Danilo Aballay}
\author[2]{Pierre Gélat}
\author[3]{Reza Haqshenas}
\author[1]{Elwin van 't Wout\thanks{Corresponding author: e.wout@uc.cl}}

\affil[1]{Institute for Mathematical and Computational Engineering, School of Engineering and Faculty of Mathematics, Pontificia Universidad Católica de Chile, Santiago, Chile}
\affil[2]{Department of Surgical Biotechnology, Division of Surgery and Interventional Science, University College London, London, United Kingdom}
\affil[3]{Department of Mechanical Engineering, University College London, London, United Kingdom}

\date{November 21, 2025}

\begin{document}

\maketitle

\begin{abstract}
Transcranial ultrasound therapy uses focused acoustic energy to induce therapeutic bioeffects in the brain. Ultrasound must be transmitted through the skull, which is highly attenuating and heterogeneous, causing beam distortion, reducing focal pressure, and shifting the target location. Computational models are frequently used to predict beam aberration, assess cranial heating, and correct the phase of ultrasound transducers. These models often rely on computed tomography (CT) images to build patient-specific geometries and estimate skull acoustic properties. However, the coarse voxel resolution of CT limits accuracy for differential equation solvers at ultrasound frequencies. This paper presents an efficient numerical method based on volume-surface integral equations to model full-wave acoustic propagation through heterogeneous skull bone. We show that our approach effectively simulates transcranial ultrasound, even when using the original CT voxels as the computational mesh, where the 0.5~mm voxel length is relatively coarse compared to the shortest wavelength of 3~mm. The method is validated against a high-resolution boundary element model using an averaged skull representation. Simulations using a CT-based skull model and a bowl transducer reveal significant beam distortion of 7.8~mm attributed to the skull's heterogeneous acoustical properties.
\end{abstract}

\section{Introduction}

Transcranial ultrasound is an emerging non-invasive modality for therapeutic applications in the brain, including for opening of the blood-brain barrier~\cite{abrahao2019, lipsman2018blood}, modulation of brain activity~\cite{tyler2018ultrasonic}, and tissue ablation~\cite{elias2016}. To ensure a successful outcome, sufficient acoustic energy must be delivered at a specific focal region in the brain. The acoustic energy at the focus may cause mechanical and thermal effects in the brain that can lead to a range of therapeutic effects. As ultrasound is transmitted through the skull, which is an attenuating and heterogeneous barrier, the transmitted energy is significantly reduced. The beam may also be distorted, causing shifts in the focal point.

Computational models of acoustic wave propagation are widely used to predict and optimize treatment outcomes~\cite{aubry2022benchmark}. These models estimate the location, shape, and intensity of the ultrasound focus, as well as the acoustic pressure levels in the bone. Computational simulations aid in transducer design, enable patient-specific planning, and are critical for developing safety guidelines for transcranial ultrasound treatments~\cite{murphy2025practical}. These numerical calculations require accurate characterizations of the propagating materials~\cite{angla2023transcranial}. Specifically, the density and speed of sound for the skull region must be retrieved from biomedical images~\cite{zadeh2025enhancing}. This process involves mapping Hounsfield units of Computed Tomography (CT) scans to acoustic parameters and interpolating voxel data onto a computational grid~\cite{drainville2023simulation}. The biomedical scans provide data on the heterogeneity and variable thickness of the skull. This information is important for assessing the skull's impact on simulation accuracy and focal aberration~\cite{li2025effects}. Therefore, the effects of these skull properties on subsequent modeling decisions must be carefully evaluated~\cite{robertson2017sensitivity}.

This manuscript will model wave propagation using the Helmholtz equation, a linear full-wave model that incorporates phase information and accounts for the scattering and transmission of ultrasound at bone-tissue interfaces~\cite{lahaye2017modern}. Using a linear model is acceptable in the context of many transcranial ultrasound applications such as neuromodulation, neurostimulation, and blood-brain barrier opening, where previous studies have shown that the nonlinear effects are negligible~\cite{mueller2017numerical, rosnitskiy2019simulation}. Furthermore, since the Helmholtz equation is a full-wave model, it provides a more accurate representation of the wave field than one-way models and ray-tracing approaches~\cite{leung2021transcranial}. 

Various numerical algorithms have been successfully applied to solve harmonic and time-dependent acoustic wave propagation problems, enabling the simulation of transcranial ultrasound. Among the most popular approaches are the pseudospectral element method~\cite{treeby2012modeling}, hybrid angular spectrum (HAS)~\cite{vyas2012ultrasound}, finite-difference time domain (FDTD)~\cite{pichardo2017viscoelastic}, and finite element methods (FEM)~\cite{marty2024transcranial}. They solve the differential equations on a computational grid covering the entire region of interest. The International Transcranial Ultrasonic Stimulation Safety and Standards (ITRUSST) conducted an intercomparison study~\cite{aubry2022benchmark}, benchmarking several numerical solvers for transcranial ultrasound and validating the accuracy of computational modeling.

In contrast to the numerical algorithms mentioned above, boundary integral methods such as the boundary element method (BEM) avoid discretizing partial differential equations altogether. Instead, they solve integral formulations for acoustic wave propagation. These approaches utilize the Green's function of the Helmholtz equation in homogeneous subdomains to derive an integral equation that is defined only at the interface between different media, thereby reducing the dimensionality of the mathematical model. Hence, the BEM naturally handles unbounded domains and does not need absorbing boundary conditions or perfectly matched layers to truncate the exterior region. Furthermore, the BEM achieves accurate results with relatively coarse meshes at high frequencies, as it avoids numerical dispersion~\cite{galkowski2023does}. In the context of therapeutic ultrasound, it has been shown that the BEM can accurately simulate the pressure field with as few as 4 elements per wavelength~\cite{haqshenas2021fast}. The OptimUS library~\cite{optimuslib} implements this approach and has been validated in benchmark studies~\cite{aubry2022benchmark}. Despite all these advantages of the BEM, the classical BEM assumes piecewise-constant material properties and cannot directly incorporate heterogeneous CT-derived skull data. Hybrid approaches, such as coupling FEM and BEM, can address this issue~\cite{shen2024efficient}, but they reintroduce the need for fine volumetric meshes and efficient solvers for high-frequency problems~\cite{wout2022fembem}. Finally, volume integral equations can model ultrasound propagation~\cite{groth2021accelerating}, even weakly nonlinear fields, but cannot solve reflections at high-contrast interfaces.

This manuscript introduces a novel computational technique for simulating transcranial ultrasound using the volume-surface integral equation (VSIE) formulation. Like BEM, VSIE is based on Green's functions, eliminating the need for artificial truncation of unbounded domains. However, VSIE overcomes key limitations of BEM as it is not limited to piecewise homogeneous geometries and can accommodate spatially varying material parameters. Furthermore, it effectively handles high-contrast interfaces and maintains low numerical dispersion throughout the simulation domain. While the mathematical properties of the VSIE method have been analyzed in the literature (e.g., \cite{costabel2015spectrum, labarca2024volume}), this is the first application (to the best of our knowledge) of this technique to simulate transcranial ultrasound. This study demonstrates VSIE's capability to simulate transcranial ultrasound at therapeutically relevant parameters (e.g., a focused field at 500~kHz) with heterogeneous bone parameters using the CT's original voxel grid. Crucially, this approach eliminates the need for interpolation or mapping of acoustic parameters between the original CT data and the computational mesh, thereby avoiding data processing errors. Finally, mesh convergence studies comparing VSIE and BEM show that uncertainties from the computational modeling design are significantly smaller than the focal aberration and pressure attenuation caused by the skull's presence.

We will show the VSIE's capacity to handle original CT data on voxel grids with a size of $0.489 \times 0.489 \times 0.5$~mm. For comparison, the wavelength in water is approximately 3 mm at a frequency of 500 kHz. Converting the Hounsfield units of the CT data to acoustic parameters yields a wavelength between 3 and 8~mm in bone, with an average of 5.3~mm. Performing an ultrasound simulation with so few grid points per wavelength typically needs dedicated solvers to avoid numerical errors. For instance, the general guideline for simulations with the $k$-Wave package, which implements a pseudospectral element method, is to use at least 8 or 10 elements per shortest wavelength~\cite{krokhmal2025comparative, robertson2017sensitivity}. However, simulations with as few as 3 elements per wavelength with the $k$-Wave library \cite{rosnitskiy2019simulation, rosnitskiy2024treatment} and FDTD simulations with 6 points per wavelength in the BabelBrain library~\cite{pichardo2023babelbrain} have also been reported in the literature. Evidently, the required number of elements per wavelength strongly depends on the specific ultrasound simulation and the design choices for the numerical method. It is worth mentioning that the numerical pollution effect of partial-differential-equation solvers causes a disproportional rise in grid sizes at increasing frequencies~\cite{babuska1997pollution}. This often necessitates dedicated GPU-based software acceleration to handle increasingly finer meshes. In contrast, integral equation solvers are less susceptible to numerical pollution effects~\cite{galkowski2023does}.

This study demonstrates that the VSIE algorithm provides accurate simulation results of transcranial ultrasound using CT-derived heterogeneous skull data. Details of the numerical formulation and CT data processing pipeline are presented in Section~\ref{sec_method}. The computational results in Section~\ref{sec_results} compare four test cases: free-field propagation, VSIE simulation with heterogeneous bone, and both VSIE and BEM simulations with homogeneous bone. Finally, mesh convergence studies confirm the numerical accuracy of our modeling strategy.

\section{Methodology}
\label{sec_method}

This section presents the computational methodology for simulating transcranial ultrasound, including the Helmholtz equation for acoustic waves, the VSIE and BEM algorithms for numerically solving the models, and the processing pipeline for converting CT data into acoustic parameters.

\subsection{Acoustic wave propagation}

We model ultrasound propagation through the skull using the Helmholtz equation, which is the most common choice for ultrasound scenarios where the acoustic field can be assumed harmonic and exhibits a linear response to the propagation materials. These are reasonable modeling assumptions for the parameters used in transcranial ultrasound~\cite{mueller2017numerical,  rosnitskiy2019simulation, aubry2022benchmark}.

\begin{figure}[!ht]
    \centering
    \includegraphics[width=0.7\linewidth]{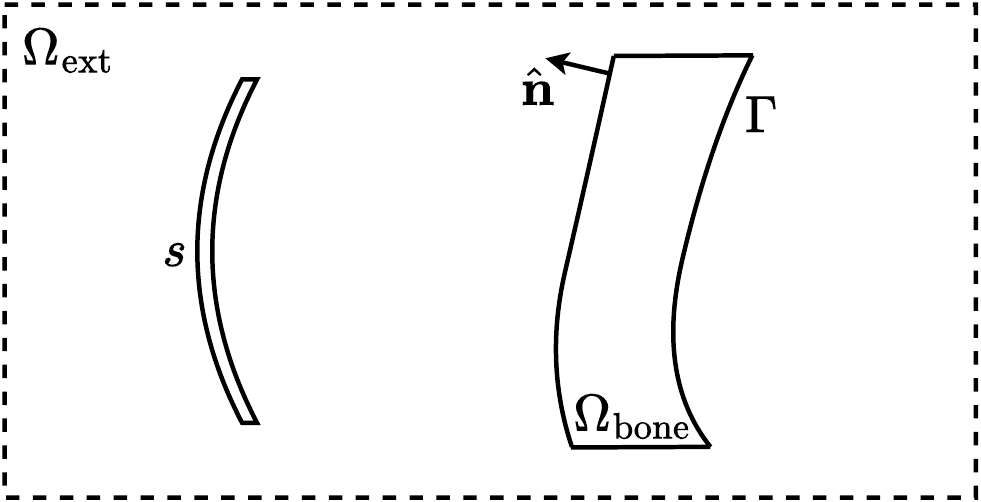}
    \caption{A sketch of the simulation domain, which consists of a transducer source~$s$, a bone region~$\bonedomain$ with normal~$\normal$ at its surface~$\Gamma$, and an unbounded exterior tissue domain~$\exteriordomain$.}
    \label{fig_sketch}
\end{figure}

Since the ultrasound field is focused, the propagation region is divided into a bounded domain, representing the part of the skull that truncates the beam, and an unbounded domain, representing soft tissue. Specifically, we denote the skull region as $\bonedomain \subset \Rthree$ and the unbounded region outside as $\exteriordomain = \Rthree \setminus \bonedomain$. The interface between these domains is denoted by~$\Gamma$, which we assume to be Lipschitz continuous with the unit normal vector~$\normal$ pointing towards~$\exteriordomain$. See Figure~\ref{fig_sketch} for a sketch of the simulation domain. The exterior region is a homogeneous domain with constant material parameters representing water. Depending on the simulation settings, the interior bone region may have heterogeneous material parameters. These modeling assumptions lead to our definitions
\begin{align}
    \rho(\mathbf{x}) = \begin{cases} \rho_0 & \text{in } \exteriordomain, \\ \rho_{1}(\mathbf{x}) & \text{in } \bonedomain; \end{cases} \label{eq_density} \\
    c(\mathbf{x}) = \begin{cases} c_0 & \text{in } \exteriordomain, \\ c_{1}(\mathbf{x}) & \text{in } \bonedomain; \end{cases} \label{eq_speed_of_sound}
\end{align}
for the material's mass density and speed of sound at location $\mathbf{x} \in \Rthree$, respectively. Mathematically, we assume differentiability of the density and continuity of the speed of sound in $\bonedomain$ for the continuous model. On a discrete level, we assume the CT data to be sufficiently smooth for the numerical evaluation of the material parameters and their gradients at each voxel center.

The source of the acoustic field, denoted by $s(\mathbf{x})$, is a weighted sum of harmonic point sources~\cite{gelat2011modelling}, each with the same angular frequency of~$\omega$. Furthermore, we assume that all sources are located in the exterior domain, i.e., $s(\mathbf{x})$ has local support in $\exteriordomain$. After extracting the harmonic time component $e^{-\imath \omega t}$, we obtain the Helmholtz equations
\begin{equation}
    \begin{cases}
        -\nabla^2 \, p(\mathbf{x}) - k_0^2 \, p(\mathbf{x}) = s(\mathbf{x}), &\mathbf{x} \in \exteriordomain; \\
        -\rho_{1}(\mathbf{x}) \,\nabla \cdot \left(\frac{1}{\rho_{1}(\mathbf{x})} \nabla p(\mathbf{x})\right) - \left(\frac{\omega}{c_{1}(\mathbf{x})}\right)^2 p(\mathbf{x}) = 0, &\mathbf{x} \in \bonedomain.
    \end{cases}
    \label{eq_helmholtz_subdomains}
\end{equation}
Here, $p(\mathbf{x})$ denotes the unknown pressure field, and $k_0 = \omega/c_0$ the constant wavenumber in the exterior domain. Since sharp jumps in material parameters are expected at the bone-tissue interface, we impose transmission conditions for continuity of the pressure field and the normal particle velocity at $\Gamma$. Precisely,
\begin{align}
    \label{eq_transmission_pressure}
    p(\mathbf{x}^+) &= p(\mathbf{x}^-) & \text{at } \Gamma; \\
    \label{eq_transmission_velocity}
    \frac1{\rho_0} \nabla p(\mathbf{x}^+) \cdot \normal &= \frac1{\rho_{1}(\mathbf{x}^-)} \nabla p(\mathbf{x}^-) \cdot \normal & \text{at } \Gamma;
\end{align}
where $\mathbf{x}^+$ and $\mathbf{x}^-$ mean the limit values from the exterior or interior domains, respectively. Furthermore, the reflected acoustic waves must propagate away from the region of interest in the far-field limit. In mathematical terms, the Sommerfeld radiation condition
\begin{equation}
    \lim_{\lvert\mathbf{x}\rvert \to \infty} \lvert\mathbf{x}\rvert \left(\nabla p(\mathbf{x}) \cdot \lvert\mathbf{x}\rvert - \imath k_0 p(\mathbf{x})\right) = 0
\end{equation}
guarantees a unique outgoing solution of the Helmholtz equation. 

\subsection{Numerical algorithm}

The Helmholtz transmission problem will be solved numerically with the VSIE and the BEM. Both numerical algorithms use an integral equation based on the fundamental solution of the Helmholtz equation. That is, both use the Green's function of the three-dimensional Helmholtz equation, given by
\begin{equation}
    G_k(\mathbf{x}, \mathbf{y}) = \frac{e^{\imath k \lvert\mathbf{x} - \mathbf{y}\rvert}}{4\pi \lvert\mathbf{x} - \mathbf{y}\rvert} \quad \text{for} \quad \mathbf{x} \ne \mathbf{y},
\end{equation}
which is valid for free-field propagation with constant wavenumber~$k$. Since the exterior domain has constant material parameters representing water, the Green's function $G_{k_0}$ models the acoustic propagation in the exterior domain. The interior Green's function can only be used in the bone regions for the simplified case of homogeneous bone parameters. Then, interior and exterior surface integral operators can be coupled at the bone-tissue interface to achieve a consistent set of boundary integral equations that can be solved with the BEM. However, this study targets the more general case where the bone's parameters are heterogeneous and derived from CT data. In that case, the VSIE method must be adopted. The main idea behind the VSIE is to write the acoustic model as a perturbed Helmholtz equation. That is, the exterior Helmholtz equation is subtracted from the system, yielding a localized model in the interior region. Applying the representation theorem to the heterogeneous Helmholtz equation provides a set of coupled volume integrals in the bone domain and surface integrals at the high-contrast interface between bone and tissue. The full derivations of the VSIE and BEM are presented elsewhere~\cite{costabel2015spectrum, groth2021accelerating, wout2021benchmarking}. Here, we summarize the final equations. The VSIE solves a system of equations involving volume and surface integrals, given by

\begin{equation}
    \begin{bmatrix}
        \ID_{\Gamma} + \DL_{\Gamma,\Gamma} & -\SL_{\Gamma,\bonedomain} + \AD_{\Gamma,\bonedomain} \\
         \DL_{\bonedomain,\Gamma} & \ID_{\bonedomain} - \SL_{\bonedomain,\bonedomain} + \AD_{\bonedomain,\bonedomain}
    \end{bmatrix}
    \begin{bmatrix}
        p_{\Gamma} \\
        p_{\bonedomain}
    \end{bmatrix}
    =
    \begin{bmatrix}
        f_{\Gamma} \\
        f_{\bonedomain}
    \end{bmatrix},
    \label{eq_vsie}
\end{equation}
where 
\begin{align}
    \ID_{\Sigma}\left[\phi\right](\mathbf{x}) &= \frac{\rho_0}{\rho(\mathbf{x})} \phi(\mathbf{x}),\\
    \SL_{\Sigma,\bonedomain}\left[\phi\right](\mathbf{x}) &= \iiint_{\bonedomain} G_{k_0}(\mathbf{x},\mathbf{y}) \frac{\rho_0}{\rho(\mathbf{y})} \left(\left(\frac{\omega}{c(\mathbf{y})}\right)^2 - \left(\frac{\omega}{c_0}\right)^2\right) \phi(\mathbf{y}) \dy,\\
    \DL_{\Sigma,\Gamma}\left[\phi\right](\mathbf{x}) &= \iint_{\Gamma} \left(\frac{\partial}{\partial \normal(\mathbf{y})} G_{k_0}(\mathbf{x},\mathbf{y})\right) \left(\frac{\rho_0}{\rho(\mathbf{y})} - 1\right) \phi(\mathbf{y}) \dy, \\
    \AD_{\Sigma,\bonedomain}\left[\phi\right](\mathbf{x}) &= \nabla_{\mathbf{x}} \cdot \iiint_{\bonedomain} G_{k_0}(\mathbf{x},\mathbf{y}) \, \nabla_{\mathbf{y}} \left( \frac{\rho_0}{\rho(\mathbf{y})}\right) \phi(\mathbf{y}) \dy,
\end{align}
for $\mathbf{x} \in \Sigma$ and $\Sigma\in\{\Gamma,\bonedomain\}$, denote the scaled mass, single-layer, double-layer, and adjoint double-layer integral operators, respectively. Furthermore,

\begin{align}
    f_{\Sigma}(\mathbf{x}) 
    &= \iiint_{\exteriordomain}G_{k_0}(\mathbf{x}, \mathbf{y}) s(\mathbf{y})\dy
    , \quad \mathbf{x}\in\Sigma
\end{align}
denotes the incident wave field. The unknown functions~$p_{\Gamma}$ and $p_{\bonedomain}$ are the pressure levels at the material interface and inside the bone region.

The VSIE algorithm is valid for both homogeneous and heterogeneous bone scenarios. In contrast, the BEM only works for homogeneous subdomains. Hence, when $\rho(\mathbf{x}) = \rho_1$ is constant in the entire bone region~$\bonedomain$, we can also solve the Helmholtz transmission problem with the BEM. Specifically, the BEM rewrites the set of homogeneous Helmholtz equations into a system of surface integral equations at the material interface. Among the many options to design a boundary integral equation (cf.~\cite{wout2021benchmarking}), we choose the Poggio–Miller–Chang–Harrington–Wu–Tsai (PMCHWT) formulation~\cite{poggio1973integral, chang1974surface, wu1977scattering}, which tends to be computationally stable and efficient for high-frequency acoustics~\cite{wout2022pmchwt}. Precisely, the BEM solves
\begin{align}
    \begin{bmatrix}
        -\DL_0 - \DL_1 & \SL_0 + \frac{\rho_1}{\rho_0} \SL_1 \\
        \HS_0 + \frac{\rho_0}{\rho_1} \HS_1 & \AD_0 + \AD_1
    \end{bmatrix}
    \begin{bmatrix}
        p_{\Gamma} \\ \partial_n p_{\Gamma}
    \end{bmatrix}
    =
    \begin{bmatrix}
        f_{\Gamma} \\ \partial_n f_{\Gamma}
    \end{bmatrix}
    \label{eq_pmchwt}
\end{align}
for the pressure and its normal gradient at the water-bone interface. Here, the operators
\begin{align}
    \SL_{j} \left[\psi\right](\mathbf{x})&=\iint_{\Gamma} G_{k_j}(\mathbf{x}, \mathbf{y}) \psi(\mathbf{y}) \dy, && \mathbf{x}\in\Gamma; \\
    \DL_{j} \left[\phi\right](\mathbf{x})&=\iint_{\Gamma} \frac{\partial}{\partial \hat{\mathbf{n}}(\mathbf{y})} G_{k_j}(\mathbf{x}, \mathbf{y}) \phi(\mathbf{y}) \dy, && \mathbf{x}\in\Gamma; \\
    \AD_{j} \left[\psi\right](\mathbf{x})&=\frac{\partial}{\partial \hat{\mathbf{n}}(\mathbf{x})} \iint_{\Gamma} G_{k_j}(\mathbf{x}, \mathbf{y}) \psi(\mathbf{y}) \dy, && \mathbf{x}\in\Gamma; \\
    \HS_{j} \left[\phi\right](\mathbf{x})&=-\frac{\partial}{\partial \hat{\mathbf{n}}(\mathbf{x})} \iint_{\Gamma} \frac{\partial}{\partial \hat{\mathbf{n}}(\mathbf{y})} G_{k_j}(\mathbf{x}, \mathbf{y}) \phi(\mathbf{y}) \dy, && \mathbf{x}\in\Gamma
\end{align}
denote the single-layer, double-layer, adjoint double-layer, and hypersingular boundary integral operators with wavenumber $k_j$, respectively~\cite{wout2021benchmarking}.

The model equations~\eqref{eq_vsie} and~\eqref{eq_pmchwt} are written as continuous integral equations. To obtain a discrete set of linear equations, we employ the VSIE and BEM numerical methods for their respective formulations. Specifically, the VSIE utilizes collocation and piecewise-constant elements on a voxel mesh. The BEM employs a Galerkin discretization of its weak formulation, with piecewise-linear elements on a triangular surface mesh. The dense linear systems arising from the VSIE and BEM algorithms were solved using the iterative Generalized Minimal Residual (GMRes) algorithm with a tolerance of~$10^{-5}$.

\subsection{Data processing}
\label{sec_data}

One of the key strengths of the VSIE algorithm is that it works for high-contrast interfaces and heterogeneous material parameters provided on a voxel grid with relatively few elements per wavelength. These numerical characteristics are essential for transcranial ultrasound simulations since material data typically come from medical images stored in pixelized slices at equidistant spacings.

Here, we use a publicly available CT scan of a human head from the MorphoSource repository \cite{morphosource}, which references the Visible Human project \cite{ratiu2003visible}. The dataset consists of 463 grayscale images in the transverse plane, with a spatial resolution of 0.489~mm, 0.489~mm, and 0.500~mm in the Cartesian $x$, $y$, and $z$ directions, respectively. Each image comprises 512 pixels in the $x$ and $y$ directions.

\begin{figure}[!ht]
    \centering
    \includegraphics[width=0.9\linewidth]{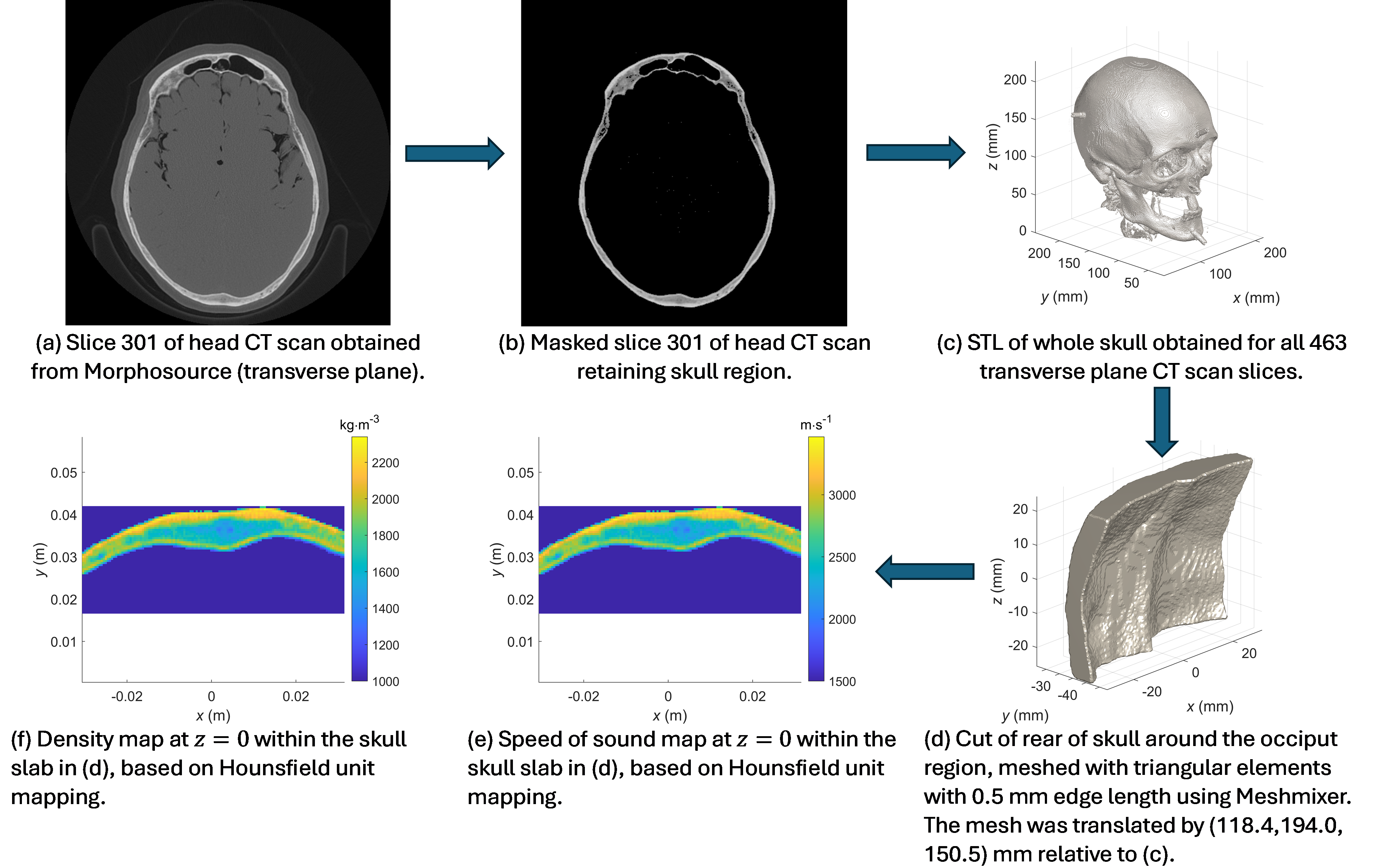}
    \caption{A visual interpretation of the algorithmic pipeline to calculate density and sound speed values inside skull bone from CT images.}
    \label{fig_pipeline}
\end{figure}

We created an algorithmic pipeline to calculate the acoustic parameters of the skull from CT images, as visualized in Figure~\ref{fig_pipeline}. A mask was initially applied for segmentation between the skull bone and other anatomical regions (skin, fat, and brain), retaining trabecular and cancellous bone. We generated 463 masked slices in the sagittal plane and an STL file representing the whole skull. Using Autodesk Meshmixer v3.5, a slab of 621 $\times$ 254 $\times$ 490 mm was cut towards the occiput region, to generate a skull slab analogous to that of benchmark~7 described in the ITRUSST intercomparison exercise~\cite{aubry2022benchmark}. These dimensions were chosen to be sufficiently large to capture the beam of the bowl transducer used for the analysis. The skull slab was then meshed with three-noded triangular elements using Meshmixer and converted from STL format to Gmsh v4.13.1 format. Since the element edge length is a short 0.25~mm, this surface mesh allows for high-accuracy BEM simulations.

The VSIE simulations work with a voxel mesh that may have different edge lengths in different directions. For this purpose, we generated a volumetric Cartesian grid that encompasses the ranges of $x$, $y$, and $z$ values defined by the bounding box of the skull slab, with the discretization determined by the location of each pixel centroid in the transverse planes and by the height $z$ corresponding to each slice. Using the solid angle method, as implemented in OptimUS \cite{optimuslib}, the grid points inside the closed surface defined by the skull slab mesh were then determined. The Hounsfield unit ($HU$) associated with each pixel for each transverse plane slice was then extracted, and the corresponding density and speed of sound were calculated as
\begin{align}
    \rho_{1}(\mathbf{x}) &= \rho_{\textrm{min}} + (\rho_\textrm{max} - \rho_\textrm{min})\frac{HU(\mathbf{x}) - HU_\textrm{min}}{HU_\textrm{max} - HU_\textrm{min}},
    \label{eq_density_HU} \\
    c_{1}(\mathbf{x}) &= c_\textrm{min} + (c_\textrm{max} - c_\textrm{min})\frac{\rho_{1}(\mathbf{x}) - \rho_\textrm{min}}{\rho_\textrm{max} - \rho_\textrm{min}}
    \label{eq_sound_speed_HU}
\end{align}
following standard procedures based on laboratory data~\cite{angla2024improved, marsac2017ex}. Here, $\rho_\mathrm{min}$ and $\rho_\mathrm{max}$ are determined from the minimum and maximum density values within the CT scan, and the same applies to $c_\mathrm{min}$ and $c_\mathrm{max}$. These extrema commonly refer to the acoustic properties of water and trabecular bone, for the least and most dense media visible in a CT scan, respectively. There is currently a lack of consensus in the scientific literature on the upper limit of these ranges~\cite{angla2024improved}, so the same values as those adopted by~\cite{marsac2017ex} were chosen. We therefore use: $\rho_\textrm{min} = 1000 \, \textrm{kg} \cdot \textrm{m}^{-3}$, $\rho_\textrm{max} = 2700 \, \textrm{kg} \cdot \textrm{m}^{-3}$, $c_\textrm{min} = 1480 \, \textrm{m} \cdot \textrm{s}^{-1}$ and $c_\textrm{max} = 4000 \, \textrm{m} \cdot \textrm{s}^{-1}$. The minimum Hounsfield value $HU_\textrm{min}$ was determined by averaging out the Hounsfield units over a region of brain from the raw CT scans, more specifically from slice 229 in the MorphoSource dataset between pixels 195 and 300 in the $x$-direction and pixels 350 and 415 along the $y$-direction. The maximum $HU_\textrm{max}$ was obtained by identifying the maximum Hounsfield unit in the raw CT scan throughout all slices. This procedure yielded $HU_\textrm{min} = 19857$ and $HU_\textrm{max} = 65535$.

\begin{figure}[!ht]
    \centering
    \includegraphics[width=0.9\linewidth]{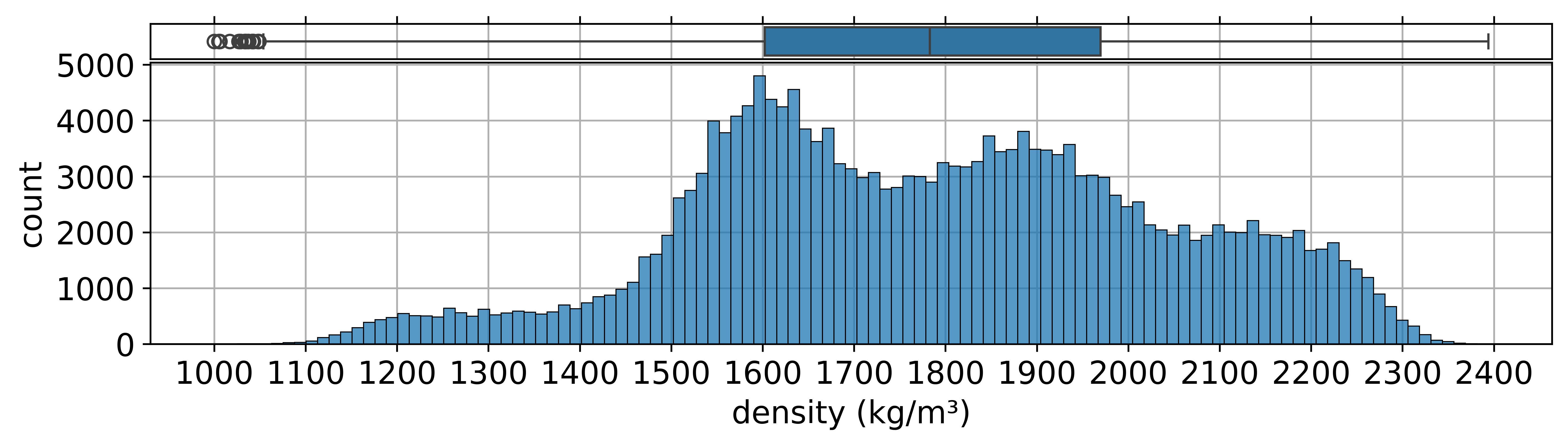}
    \includegraphics[width=0.9\linewidth]{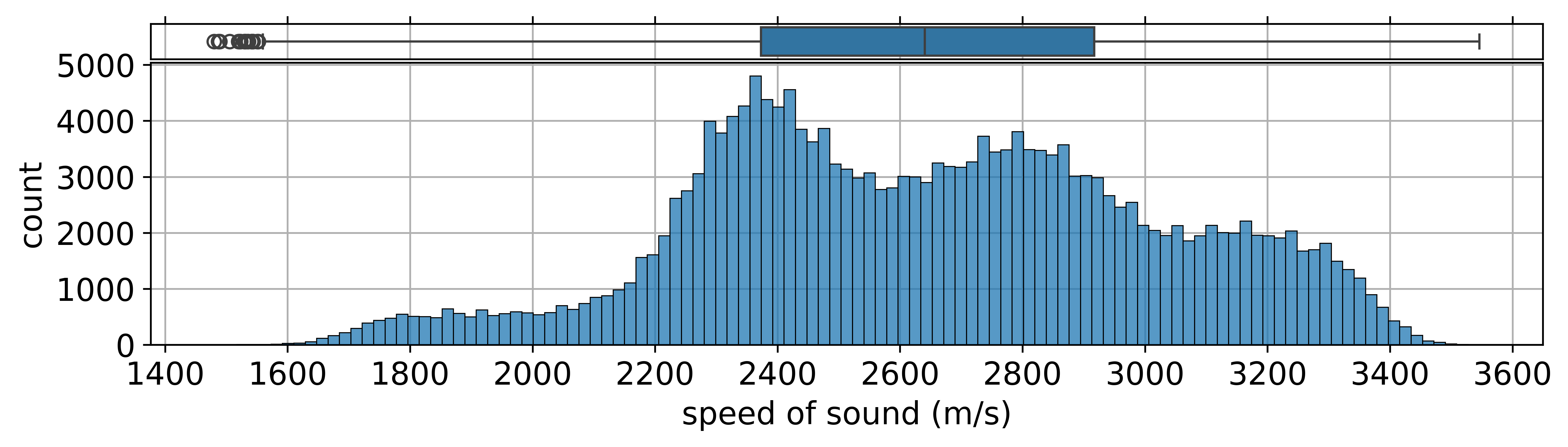}
    \caption{The histograms and boxplots show the distribution of the voxel values concerning the density and speed of sound. The computational grid of the skull slab consists of 199,693 voxels.}
    \label{fig_histograms}
\end{figure}

The pipeline of this procedure to extract acoustic parameters from CT images is summarized in Figure~\ref{fig_pipeline}. The final product is a collection of 199,693 voxels inside the skull slab, each with a value for the density and speed of sound directly calculated from the Hounsfield maps. Figure~\ref{fig_histograms} displays the distribution of the acoustic values across the skull slab, which the VSIE algorithm will use to simulate ultrasound propagation through the heterogeneous bone.

For computational models that use homogeneous material parameters for the bone region, the density and speed of sound of the skull slab are taken as the arithmetic mean of all voxel data. We obtain 1788.0~kg/m$^3$ and 2648.1~m/s for the mean density and speed of sound, respectively. More elaborate homogenization approaches could be adopted, but taking the average has been shown to provide sufficiently accurate results for acoustic wave propagation through skull bone (see, e.g.,~\cite{angla2024new}).

\subsection{Ultrasound transducer}

\begin{figure}[!ht]
    \centering
    \includegraphics[width=0.8\linewidth]{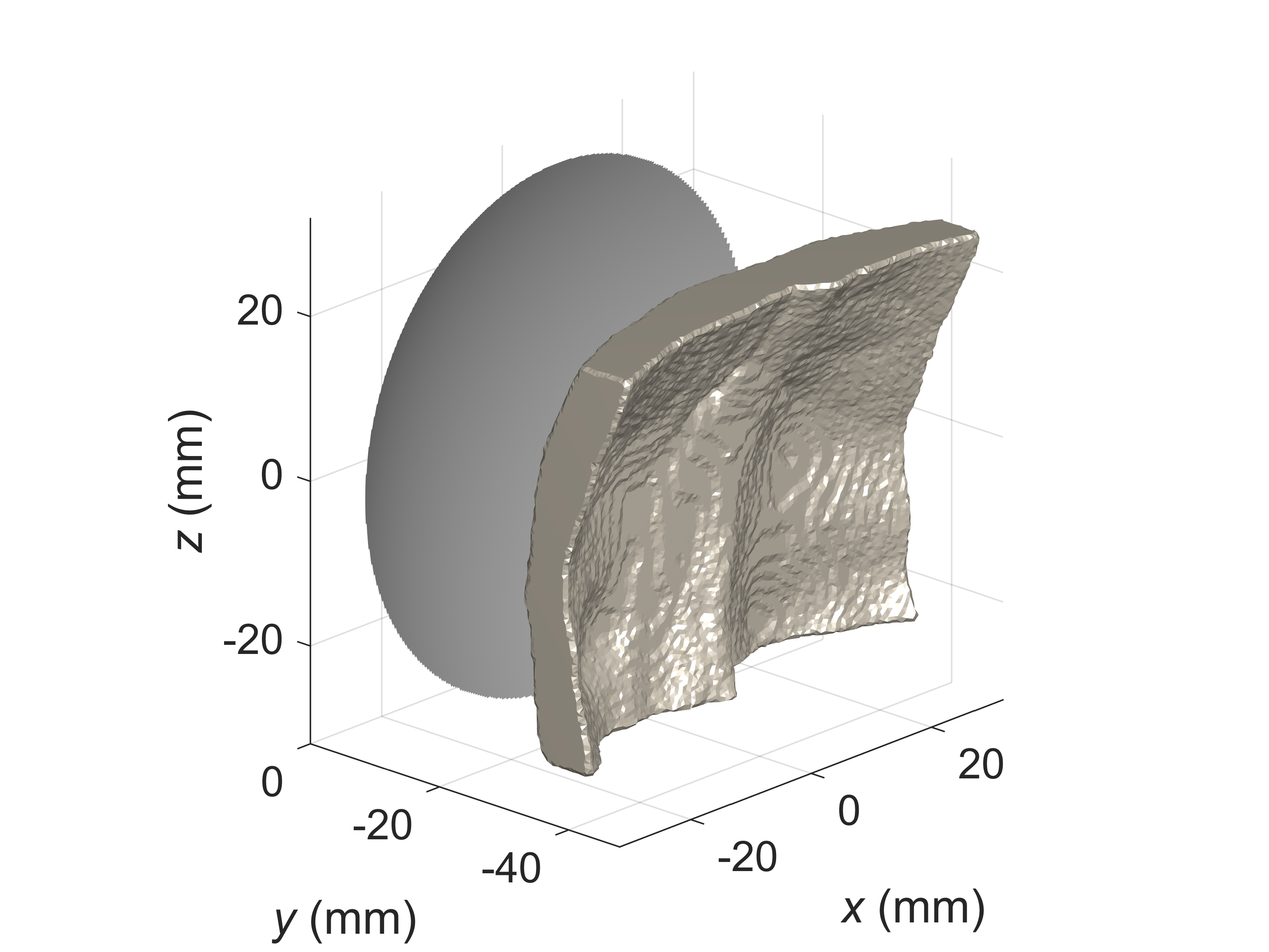}
    \caption{The geometry of the ultrasound transducer and the skull slab.}
    \label{fig_skull_transducer}
\end{figure}

A single-element bowl transducer with an outer radius of 32~mm and a radius of curvature of 64~mm is used as the source. The center of the bowl is located at the global origin of the coordinate system and faces towards the negative $y$~direction. The transducer is located at least 12~mm away from the skull and 24~mm distance on the focal axis; see Figure~\ref{fig_skull_transducer}.

\section{Results}
\label{sec_results}

Both the VSIE and BEM models have been applied to simulate the propagation of focused ultrasound through the skull slab.

\subsection{Computational configurations}

We have implemented the VSIE algorithm to solve the Helmholtz equation and simulate ultrasound propagation. The implementation is in Python and uses the Numba library for shared-memory parallel computing~\cite{numba}. The BEM simulations were performed using the open-source OptimUS library (v.~0.2.1)~\cite{optimuslib}, which uses the Bempp library (v.~3.3.5) to perform the BEM calculations~\cite{smigaj2015solving}. The simulations were executed on a compute node with 32~cores shared across two Intel(R) Xeon(R) Silver 4216 2.10 GHz processors and a total of 2048~GB RAM. The heterogeneous VSIE simulation with CT-derived skull parameters used 972~GB of RAM and took 18~minutes for creating the matrix and 8:38~hours for solving the linear system.

\subsection{Geometry and material parameters}

As outlined in Section~\ref{sec_data}, the VSIE uses exactly the same voxels as the ones in the original CT scan data. Precisely, the hexahedral volume mesh of the skull slab consists of 199,693 voxels, each with a size of $\Delta x = 0.489$~mm, $\Delta y = 0.489$~mm, and $\Delta z = 0.500$~mm in the Cartesian $x$, $y$, and $z$-directions, respectively. The triangular surface mesh for the BEM at the material interface has 111,764 vertices and a characteristic mesh size of 0.25~mm. This resolution of the BEM mesh was chosen finer than the VSIE to obtain a high-accuracy reference solution.

The bowl transducer emits an ultrasound field with a frequency of 500~kHz and a piston velocity of 0.04~m/s. For the material parameters in the exterior domain, we use default values of water, i.e., a wave speed of 1500~m/s and a density of 1000~kg/m$^3$. Although all models allow for complex wavenumbers, no attenuation was included in the model for this study.

For these parameters, the wavelength in the exterior water region is 3~mm and the wavelength in the homogeneous skull is 5.3~mm. Compared to the characteristic grid size of 0.489~mm in the voxel mesh, we have at least 6 elements per wavelength for the VSIE. The BEM uses a 0.25~mm resolution for the triangular surface mesh, yielding at least 12 elements per wavelength.

\subsection{Transcranial ultrasound simulations}

We simulated the transcranial ultrasound propagation with the VSIE and BEM algorithms on the skull slab, as described in Section~\ref{sec_method}. The primary purposes of the computational simulations are to quantify the focal aberrations caused by bone heterogeneity and to investigate the impact of different modeling approaches on the focusing capacity of transcranial ultrasound fields. Specifically, we compare four computational approaches.
\begin{enumerate}
    \item The incident wave field from the bowl transducer with free-field propagation in water. This case serves as a benchmark for numerical simulations, allowing focal aberrations due to bone presence to be measured.
    \item The simulated pressure field with the VSIE applied to heterogeneous voxel data taken directly from the CT dataset. That is, the VSIE voxels coincide with the voxels from the biomedical images, and the values of the density and speed of sound come from the Hounsfield unit maps.
    \item The simulated pressure field with the VSIE applied to averaged homogeneous voxel data. Precisely, it uses the original voxel locations but with constant values for density and speed of sound taken as the mean of the Hounsfield values across the skull slab.
    \item The simulated pressure field with the BEM applied to a smooth surface mesh at the bone-tissue interface. The free-field Green's functions in the BEM take the same averaged values for density and speed of sound as the VSIE.
\end{enumerate}
Each of the computational approaches allows for calculating the complex-valued pressure at any location in the three-dimensional coordinate system. We calculate them in the voxel centroids inside the bone and on several slices through the region of interest.

\subsection{Focal aberrations of the acoustic fields}

\begin{figure}[!ht] 
    \centering
    \includegraphics[width=\linewidth]{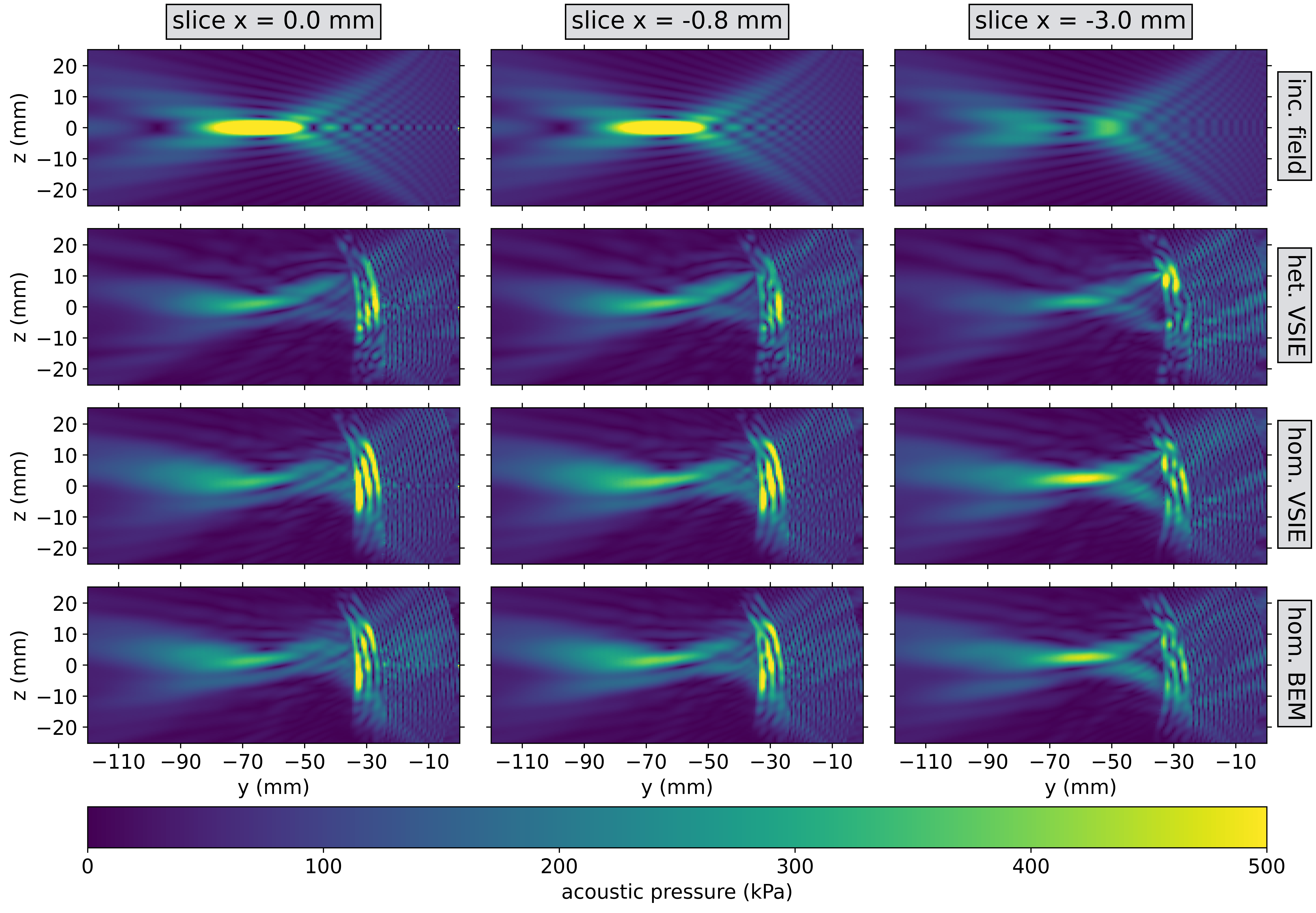}
    \caption{The magnitude of the acoustic field for the four cases: the incident field emitted by the bowl transducer (top row), the total field simulated with the VSIE and heterogeneous bone (second row), the VSIE with homogeneous bone (third row), and the BEM simulation with homogeneous bone (bottom row). The pressure field is calculated on different slices on the $x$-axis. The center of the bowl transducer is located at the global origin and emits an acoustic field of 500~kHz in the negative $y$-direction.}
    \label{fig_fields}
\end{figure}

\begin{figure}[!ht] 
    \centering
    \includegraphics[width=\linewidth]{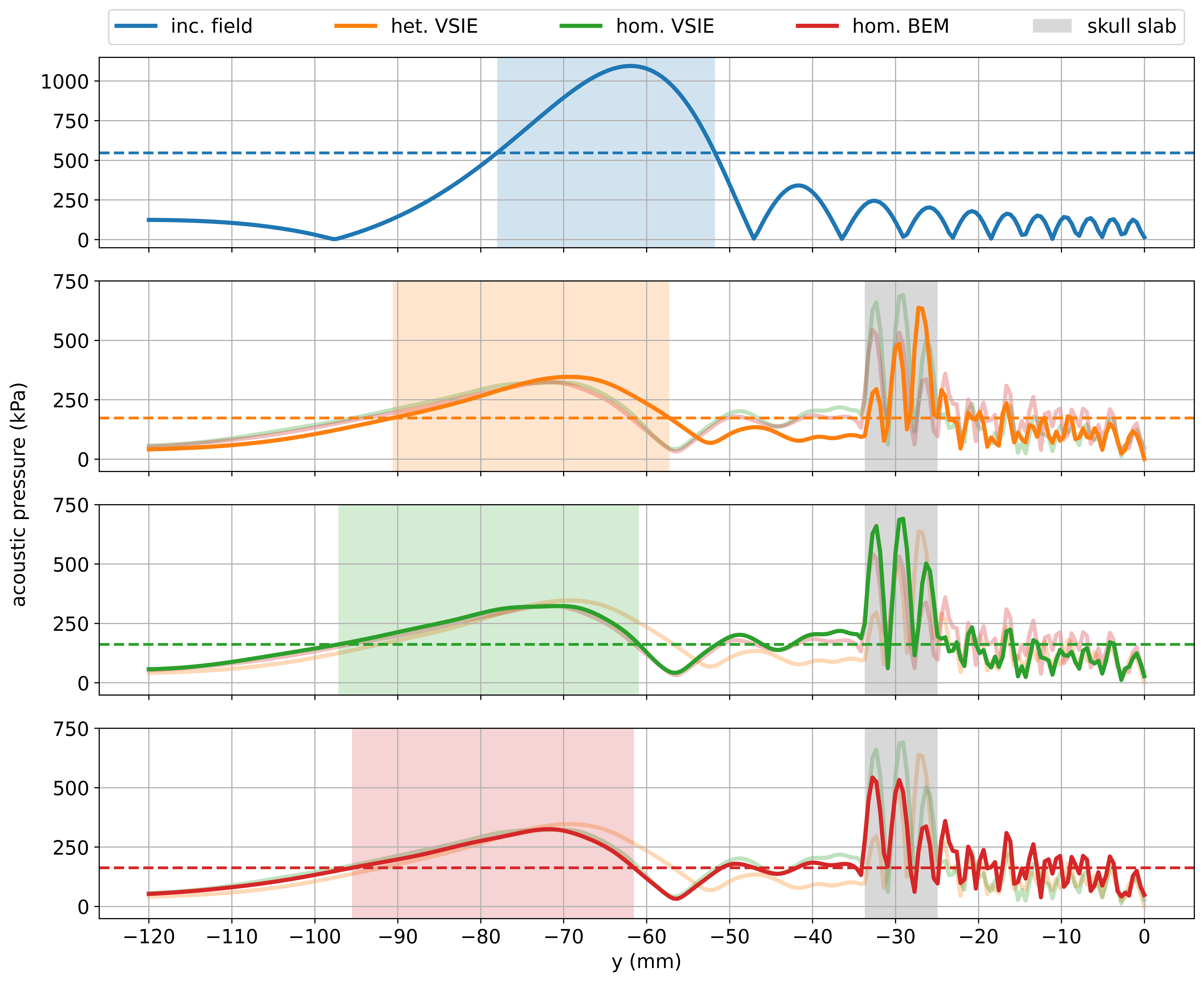}
    \caption{The acoustic pressure levels on the propagation axis ($x = 0$~mm and $z = 0$~mm), with the transducer located in the origin, emitting in the negative $y$-direction. The top panel depicts the incident wave field, and the other three highlight the differences between the simulations with the heterogeneous VSIE, homogeneous VSIE, and homogeneous BEM. The location of the skull slab is depicted in grey. The horizontal lines and the shaded areas depict the full width at half maximum (FWHM) pressure level.}
    \label{fig_fwhm}
\end{figure}

The acoustic pressure fields for these four simulations are depicted in Figure~\ref{fig_fields} on the three slices $x = 0.0$~mm, $x = -0.8$~mm, and $x = -3.0$~mm. Figure~\ref{fig_fwhm} displays the pressure levels on the propagation axis, i.e., $x = 0$~mm and $z = 0$~mm.

The pressure maps clearly display the reduced energy at the focus and the aberration of the focal maximum due to the presence of the bone in the ultrasound beam path. Table~\ref{table_statistics_focus} presents various statistics of the focal aberration calculated using different modeling approaches, including $p_\mathrm{max}$, the maximum pressure amplitude at the focus, and its corresponding location.

\begin{table}[!ht]
    \centering
    \caption{The pressure level and location of the peak pressure at the focus, calculated for the incident field and the pressure fields from the different simulation approaches.}
    \label{table_statistics_focus}
    \begin{tabular}{l|rr}
    simulation         & $p_\mathrm{max}$ (kPa) & focus location (mm) \\ \hline
    incident field     & 1094.3     & (0.00, -61.85, 0.00) \\
    heterogeneous VSIE & 397.6      & (-0.80, -65.54, 1.01) \\
    homogeneous VSIE   & 518.8      & (-3.00, -58.15, 2.27) \\
    homogeneous BEM    & 477.0      & (-2.70, -59.08, 2.52) \\
    \end{tabular}
\end{table}

\subsection{Acoustic pressure levels in the skull slab}

\begin{figure}[!ht] 
    \centering
    \includegraphics[width=\linewidth]{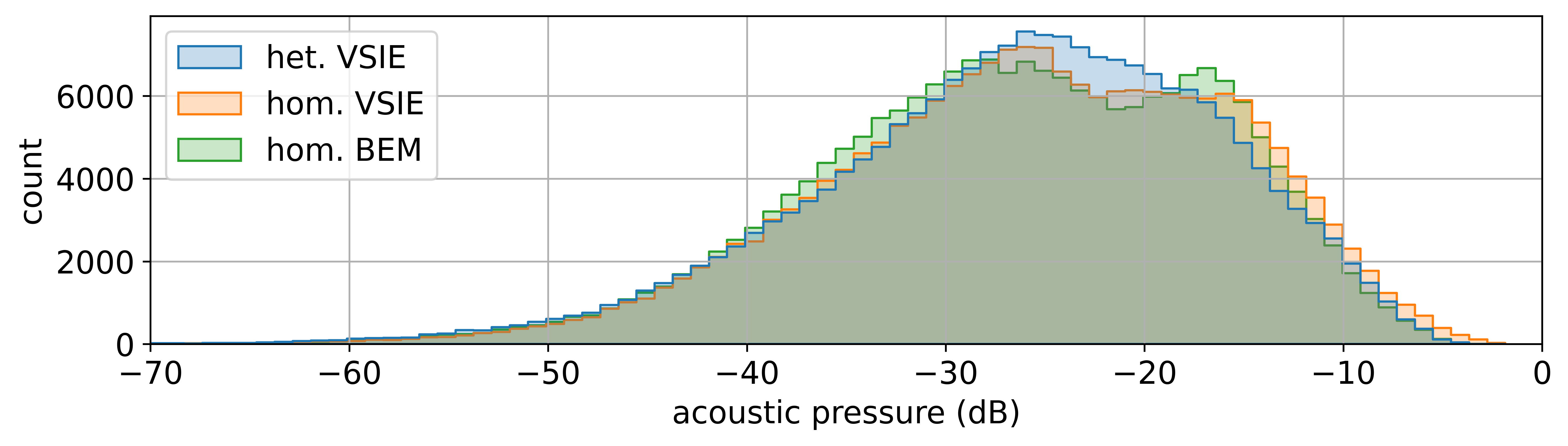}
    \caption{Histograms of the magnitude of the acoustic pressure levels calculated in each of the 199,693 voxels of the skull slab. The reference pressure for the dB scale is the peak pressure of the incident field in the focus. The colors represent the three simulations: VSIE for heterogeneous bone (blue), VSIE for homogeneous bone (red), and BEM for homogeneous bone (green).}
    \label{fig_hist_bone}
\end{figure}

A common challenge in transcranial ultrasound therapies is avoiding overheating of the bone, which can be damaging. High pressure levels are expected since the ultrasound must be transmitted through the skull. Figure~\ref{fig_hist_bone} provides a histogram of pressure levels in the skull slab, and Table~\ref{table_statistics_bone} summarizes this information in several metrics. Specifically, we use the spatial root mean square (RMS) to quantify the average pressure levels in the skull slab~\cite{gelat2025evaluation}. This metric depends on the L2-norm and is given by
\begin{equation}
    p_\mathrm{RMS} = \sqrt{\frac1N \sum_{i=1}^N \lvert p_i \rvert^2}
\end{equation}
where $p_i$ denotes the pressure level at the center of each of the $N =$ 199,693 voxels. Since the acoustic energy is localized, the average pressure levels may provide biased outcomes as a metric for cranial heating. Hence, we also calculate the 90th percentile value (P90) of the magnitude of the complex-valued pressure field in the skull slab, denoted by $p_\mathrm{P90}$. Finally, we convert the metrics from Pa to dB as
\begin{equation}
    L_\mathrm{dB} = 20\log_{10}\left(\frac{L_\mathrm{Pa}}{\lvert p_0 \rvert}\right),
    \label{eq_pa_to_db}
\end{equation}
where the reference value $p_0$ was chosen as the pressure of the incident field in the geometric focus, which equals 1094.3~kPa.

\begin{table}[!ht]
    \centering
    \caption{Statistics on the acoustic field inside the bone for the different simulation approaches. The RMS and P90 metrics are calculated over all voxels in the skull slab.}
    \label{table_statistics_bone}
    \begin{tabular}{l|rrrr}
    simulation & $p_\mathrm{RMS}$ (kPa) & $p_\mathrm{RMS}$ (dB) & $p_\mathrm{P90}$ (kPa) & $p_\mathrm{P90}$ (dB) \\ \hline
    heterogeneous VSIE & 129.77 & -18.52 & 215.51 & -14.11 \\
    homogeneous VSIE & 145.40 & -17.53 & 241.06 & -13.14 \\
    homogeneous BEM & 129.87 & -18.51 & 218.10  & -14.01 \\
    \end{tabular}
\end{table}

\subsection{Mesh convergence study}

The computational results of transcranial ultrasound focusing were achieved by applying the VSIE directly to the CT voxel data, which has a resolution of $0.489 \times 0.489 \times 0.5$~mm. To confirm that the numerical algorithms achieve working precision at this grid resolution, we also performed mesh convergence studies. We consider a piecewise homogeneous material with the same averaged values for bone as in Section~\ref{sec_data} to avoid introducing interpolation errors in the material parameters at different grid resolutions.

We performed two sets of benchmarks. The first benchmark considers a rectangular box that represents the skull slab as closely as possible. Precisely, the box has the same extent and volume as the skull slab, resulting in a thickness of approximately 7.8~mm. The location of the geometry was determined by centering it along the $y$-direction in the bounding box of the skull slab. The second benchmark uses meshes that approximate the CT-segmented skull slab. For this purpose, we created surface meshes at the boundary of the skull slab at different resolutions. The voxel meshes at different resolutions were created by detecting if the voxel's center is located inside the enclosed surface mesh. All simulations were performed at 500~kHz, where the wavelengths in the water and bone regions are approximately 3~mm and 5.3~mm, respectively.

\begin{figure}[!ht] 
    \centering
    \includegraphics[width=\linewidth]{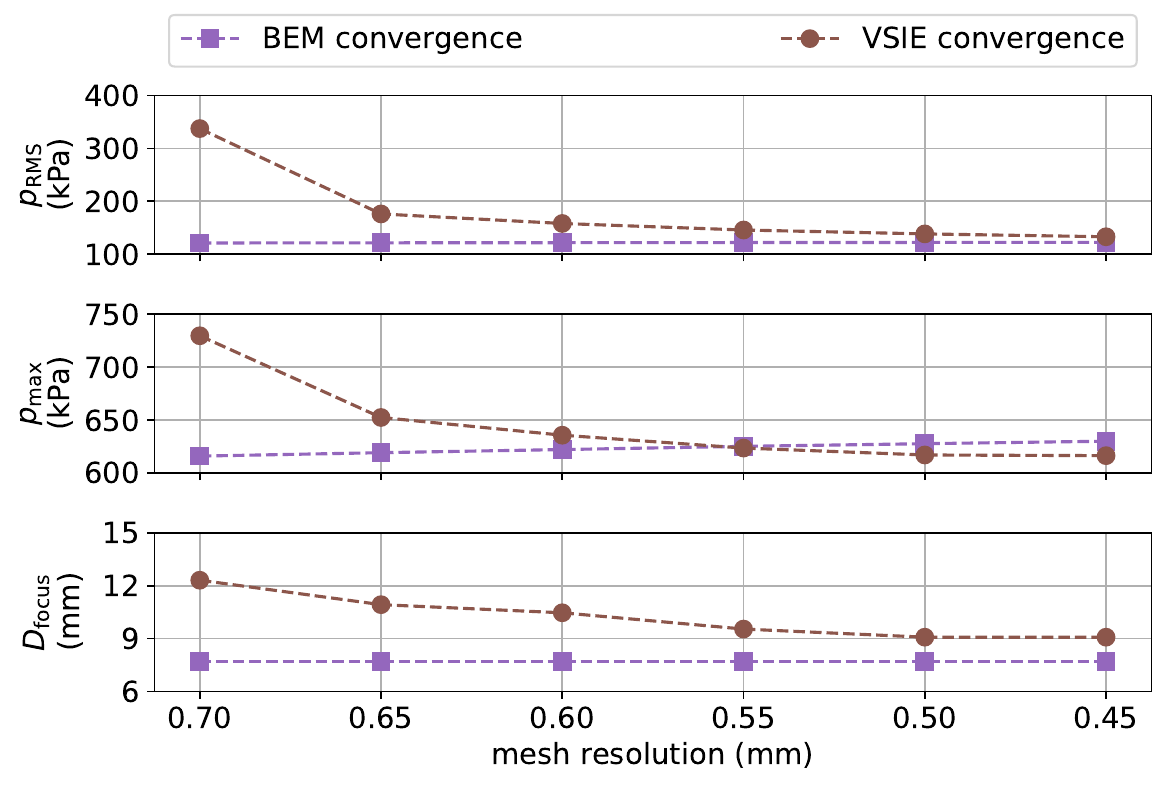}
    \caption{The mesh convergence of the VSIE and BEM for a homogeneous rectangular box.}
    \label{fig_convergence_box}
\end{figure}

Figure~\ref{fig_convergence_box} presents the convergence metrics for the rectangular box, i.e., the location and pressure at the focal point and the average pressure inside the skull. Here, $D_\mathrm{focus}$ denotes the distance between the location of the peak pressure and the geometric focus of the transducer. The results clearly show mesh convergence and consistency between the VSIE and BEM numerical algorithms. For example, at 0.5~mm mesh resolution, the relative pressure differences are only 13.2\% for the average level in the bone and 1.7\% at the focal peak. The difference in focal location is only 1.4~mm. The relative differences remain small at the finest mesh with 0.45~mm resolution, which is finer than the CT voxels. Hence, both VSIE and BEM algorithms yield accurate results at mesh resolutions comparable to those of CT data. Furthermore, the quicker convergence of the BEM compared to the VSIE is also expected, given that our implementations use more accurate discretization for the BEM. That is, P0-collocation on a voxel grid for the VSIE, versus P1-Galerkin on triangular surface meshes for the BEM.

\begin{figure}[!ht] 
    \centering
    \includegraphics[width=\linewidth]{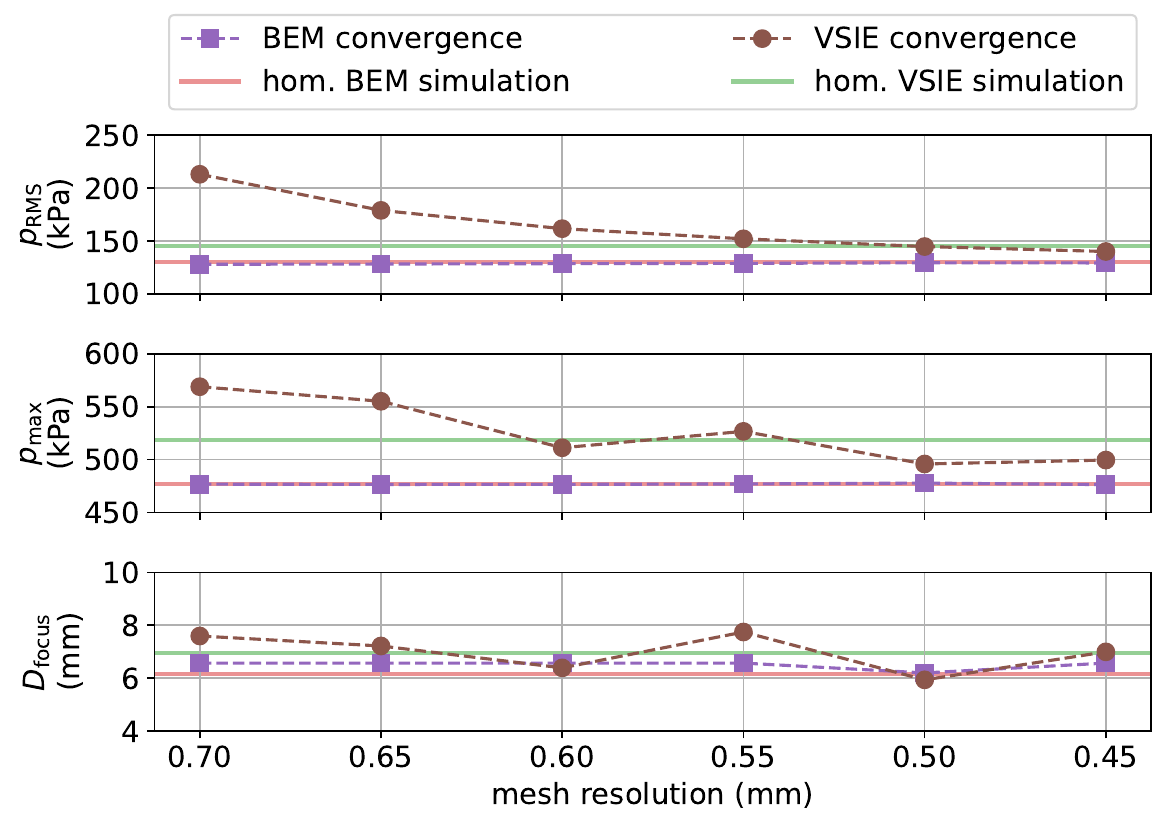}
    \caption{The mesh convergence of the VSIE and BEM for a homogeneous skull slab. The horizontal lines depict the values for the homogeneous simulations on the original CT voxels, as provided in Tables~\ref{table_statistics_focus} and~\ref{table_statistics_bone}.}
    \label{fig_convergence_skull}
\end{figure}

Figure~\ref{fig_convergence_skull} presents the convergence study on various meshes for the skull slab geometry. Again, we see a consistent convergence pattern for both numerical algorithms, with the BEM being more accurate thanks to its linear elements, compared to the VSIE's piecewise-constant voxel approximation of the pressure field. The convergence of the VSIE is slower than in the benchmark of the rectangular box, which is likely due to staircasing effects at the surface of the voxelized representation of the skull slab. The smoother convergence of the $p_\mathrm{RMS}$ compared to the metrics at the focus can be explained by the fact that the RMS is an average value over many voxels, while the $p_\mathrm{max}$ and $D_\mathrm{focus}$ depend on the peak pressure. Furthermore, the reference lines in Figure~\ref{fig_convergence_skull} depict the computational results of the homogeneous simulations on the original CT voxels, which have a resolution of $0.489 \times 0.489 \times 0.5$~mm. We observe that the differences between BEM and VSIE on the CT voxels fall within the convergence bandwidth of the algorithms.

\section{Discussion}

All computational simulations studied the acoustic propagation from an ultrasound bowl transducer and its focusing capacity when guided through a skull slab. Before discussing the effect of the bone on the acoustic field, let us consider the free-field propagation, depicted in the top row of Figure~\ref{fig_fields}. As expected, the field is highly focused around the geometric center of the bowl transducer, located at $(0,-64,0)$~mm in the coordinate system. The acoustic pressure levels exceed 1~MPa at the geometric focus, confirming constructive interference at the focus. Moving away from the geometric focus, the fields reduce in strength, as expected. Finally, the short wavelength of 3~mm in water is also clearly visible in the displayed field.

\subsection{Effect of bone heterogeneity on focused ultrasound}

When applying the VSIE to the voxel data with Hounsfield-mapped material parameters for the skull slab, the acoustic field should display features such as reflection and transmission. Indeed, the second row of Figure~\ref{fig_fields} clearly shows different acoustic fields than free-field propagation. We highlight three distinctive patterns. First, the energy delivered to the focal regions is reduced considerably due to the attenuation of the bone region. The maximum pressure at the focus is 397.6~kPa, which is only 36.3~\% of the incident field pressure. Second, the peak pressure is now located at (-0.8, -65.54, 1.01)~mm, an aberration of 3.9~mm compared to the incident field's focus. Third, the simulations reveal high acoustic pressure levels within the bone region, with a P90 pressure level of 215.5~kPa. These hotspots may compromise the safety of transcranial ultrasound treatment.

The comparisons between the incident field and the pressure levels simulated with the VSIE confirm the expected features of high pressure levels in the skull slab and focal aberrations of the ultrasound beam. A key feature of the proposed VSIE algorithm is its capacity to simulate acoustic propagation through heterogeneous materials with acoustic parameters provided on a voxel grid. This allows for analyzing the effect of bone heterogeneity on focused ultrasound by comparing heterogeneous VSIE simulations with those on mean material values.

The third row of Figure~\ref{fig_fields} depicts the acoustic field simulated by the VSIE on homogeneous material parameters. Compared to the heterogeneous VSIE simulation, the first-order characteristics of the acoustic pressure field are the same, with attenuation at the focus and hotspots inside the bone. However, a quantitative comparison does highlight differences that may impact the effectiveness and safety of focused ultrasound therapy. The most distinguishable feature is the higher peak pressure of 518.8~kPa at the focus in the homogeneous simulation compared to 397.6~kPa for the heterogeneous model (see Table~\ref{table_statistics_focus}). This overestimation of the peak pressure is consistent with the literature (cf.~\cite{angla2024improved}).

Furthermore, significant differences are visible inside the skull slab. Standing wave patterns are present, and the pressure values inside the bone are, in most regions, higher in the homogeneous model than in the heterogeneous one. The VSIE simulations also indicate a focal displacement perpendicular to the propagation axis. For example, examining the slice at $x = -3$~mm, the transcranial simulations reveal higher pressure levels near the focus than those present in the incident wave field. The geometry of the skull slab must have caused this focal aberration.

Our VSIE simulations were performed on the original CT data, thus avoiding interpolation errors between the CT voxels and the computational grid. This approach may, in theory, be better than using averaged acoustic parameters as it incorporates all the information available from CT scans to model heterogeneous bone. However, several studies that compare computational simulations with experimental data on ex vivo models indicate a more elaborate situation~\cite{yoon2018multi, krokhmal2025comparative, li2025effects}. For instance, the CT imaging process is prone to measurement errors, and the Hounsfield-unit maps are retrieved from laboratory experiments on phantoms that may be inaccurate for the skull bone~\cite{murphy2025practical}. Numerical algorithms may thus estimate the attenuation from heterogeneous bone models incorrectly due to the noise in the data. Choosing an appropriate modeling pipeline for acoustic parameters remains an active topic of academic study~\cite{angla2023transcranial}.

\subsection{Comparison between VSIE and BEM algorithms}

This study is the first to use the VSIE algorithm to simulate therapeutic ultrasound scenarios. We confirmed the validity of the numerical results with two strategies. First, the VSIE results are compared to high-accuracy BEM simulations. Second, mesh convergence studies highlight the numerical accuracy of the VSIE method.

To validate the VSIE's numerical accuracy at realistic transcranial ultrasound parameters, the VSIE results are compared with the BEM simulations for the homogeneous skull. The BEM algorithm implemented in the open-source OptimUS library has been validated against state-of-the-art software in an intercomparison exercise~\cite{aubry2022benchmark}. The mathematical model of the BEM is the closest alternative to VSIE, also using integral equations and Green's functions. However, it only works for piecewise homogeneous domains. Hence, we can only compare the BEM with the VSIE simulation on the homogeneous skull slab. The primary difference lies in the meshing strategy. While the VSIE utilizes voxels from the CT scans, the BEM employs a surface mesh at the border of the skull slab. This surface mesh is created by fitting it to segmented voxel data, a data processing step that may add modeling inaccuracies. The bottom row of Figure~\ref{fig_fields} displays similar fields to the homogeneous VSIE simulations. The focus is located at almost the same location (see Table~\ref{table_statistics_focus}), and its amplitude does not change appreciably, confirming the consistency between VSIE and BEM.

Even though the VSIE and BEM simulations on the homogeneous skull slab are consistent, the pressure values have differences up to 1~dB and a shift of up to 1~mm of the focal location between the two numerical methods. On the one hand, those differences between methodologies are much smaller than the effect of bone heterogeneity on the simulated pressure field. On the other hand, those differences are noticeable and must be carefully evaluated. For this purpose, we performed mesh convergence studies for the BEM and VSIE on meshes with increasingly fine resolution.  As expected, the numerical results in Figure~\ref{fig_convergence_skull} show that using finer meshes diminishes the algorithms' discretization error and geometric error, including the staircasing effect. Importantly, the differences between the homogeneous VSIE and BEM simulations on the CT voxels have a similar magnitude to the relative differences in the mesh convergence study. Hence, the different results between the two numerical methods can be attributed to the accumulation of numerical errors at the mesh resolution typical of CT voxel data. Moreover, the observed differences between BEM and VSIE are consistent with intercomparison exercises on transcranial reported in literature~\cite{aubry2022benchmark}. Generally speaking, the differences between the homogeneous BEM and VSIE simulations fall within the numerical accuracy margins, thus confirming the validity of our modeling approach.

\section{Conclusions}

This work presents a novel three-dimensional computational method for solving the Helmholtz equation using a VSIE formulation, specifically tailored for transcranial ultrasound simulations. The key advantage of our method is high numerical accuracy and the ability to work directly on standard CT voxel data. The numerical experiments demonstrate that accurate field computations are possible with as few as six voxels per shortest wavelength.

Furthermore, we investigated the impact of using averaged material parameters for density and speed of sound compared to the Hounsfield-unit-mapped voxel data. Our results on the considered dataset show an aberration of 7.8~mm in the location of the focal peak between the two modeling approaches and higher pressure levels within the skull bone when the heterogeneous structure is ignored. Additionally, our VSIE results showed excellent agreement with high-resolution BEM simulations for a homogeneous skull, validating the accuracy of the method.

\section*{Software and data availability}

\begin{itemize}
    \item OptimUS (open-source Python library): \href{https://github.com/optimuslib}{github.com/optimuslib}
    \item Human skull CT scan: \href{https://www.morphosource.org/concern/media/000367572}{www.morphosource.org/concern/media/000367572}
\end{itemize}

\section*{Acknowledgments}

This work was financially supported by the Agencia Nacional de Investigación y Desarrollo (ANID), Chile [FONDECYT 1230642].

\printbibliography

\end{document}